\documentclass[10pt, 3p,fleqn]{elsarticle}

\makeatletter
\def\ps@pprintTitle{%
 \let\@oddhead\@empty
 \let\@evenhead\@empty
 \def\@oddfoot{\centerline{\thepage}}%
 \let\@evenfoot\@oddfoot}
\makeatother

\usepackage[british]{babel}
\usepackage[utf8]{inputenc}
\usepackage{graphics}
\usepackage{graphicx}
\usepackage{amsmath,amssymb,esint}
\usepackage{lineno}
\usepackage{multirow}
\usepackage{subfig}
\usepackage{natbib}
\usepackage{eucal}
\usepackage{mathtools}
\usepackage{placeins}
\usepackage{enumerate}
\usepackage[usenames,dvipsnames]{xcolor}
\usepackage[colorlinks]{hyperref}
\usepackage{setspace}

\usepackage{tikz}
\usetikzlibrary{calc}
\usetikzlibrary{shapes.geometric, arrows}
\usetikzlibrary{arrows.meta}
\tikzset{>={Latex[width=1mm,length=1mm]}}

\tikzstyle{process} = [rectangle, minimum width=2.5cm, minimum height=1cm, text
centered, text width = 3cm, draw=black]
\tikzstyle{decision} = [diamond, minimum width=2.5cm, minimum height=1cm,
aspect=2,inner sep=-0.5ex,text centered, text width = 2.5cm, draw=black]
\tikzstyle{arrow} = [thick,->,>=stealth]
\tikzstyle{line} = [draw, -latex']

\biboptions{sort&compress}
\journal{Journal of Computational Physics}

\begin{document}

\begin{frontmatter}

\title{A unified algorithm for interfacial flows with incompressible and compressible fluids}

\author{Fabian Denner\corref{cor1}}
\author{Berend van Wachem}

\address{Chair of Mechanical Process Engineering, Otto-von-Guericke-Universit\"{a}t Magdeburg,\\ Universit\"atsplatz 2, 39106 Magdeburg, Germany}

\cortext[cor1]{fabian.denner@ovgu.de}

\begin{abstract}
The majority of available numerical algorithms for interfacial two-phase flows either treat both fluid phases as incompressible (constant density) or treat both phases as compressible (variable density). This presents a limitation for the prediction of many two-phase flows, such as subsonic fuel injection, as treating both phases as compressible is computationally  expensive due to the very stiff pressure-density-temperature coupling of liquids. A framework with the capability of treating one phase compressible and the other phase incompressible, therefore, has a significant potential to improve the computational performance and still capture all important physical mechanisms. We propose a  numerical algorithm that can simulate interfacial flows in all Mach number regimes, ranging from $M=0$ to $M > 1$, including interfacial flows in which compressible and incompressible fluids interact, within the same pressure-based framework and conservative finite-volume discretisation. For interfacial flows with only incompressible fluids or with only compressible fluids, the proposed pressure-based algorithm and finite-volume discretisation reduce to numerical frameworks that have already been presented in the literature. Representative test cases are used to validate the proposed algorithm, including mixed compressible-incompressible interfacial flows with acoustic waves, shock waves and rarefaction fans.
\end{abstract}

\begin{keyword}
Interfacial flows \sep All-speed flows \sep Finite-volume method \sep Volume-of-Fluid method \sep Shock capturing 
\end{keyword}

\end{frontmatter}

\section{Introduction}
\label{sec:intro}
Numerical methods and algorithms for interfacial flows have, so far, typically been developed either for the simulation of compressible fluids or the simulation of incompressible fluids, as a result of the different dominant physical mechanisms, the different mathematical characteristics of the governing equations and the different numerical challenges at different flow speeds.

While in reality all fluids are compressible, with a compressibility given as $\beta = \mathrm{d}\rho/(\rho \, \mathrm{d}p$), where $\rho$ is the density and $p$ is the pressure, a frequent assumption in modelling fluid flows is that the fluid is {\em incompressible} ($\mathrm{d}\rho=0$). From a numerical viewpoint, the distinction between incompressible and compressible fluids is important and the Mach number $M=U/a$, with $U$ the flow speed and $a$ the speed of sound, is of particular importance, as it determines the mathematical nature of the governing conservation laws. For {\em incompressible fluids} the density is constant along the fluid particle trajectories and the speed of sound is $a \rightarrow \infty$, with $M \rightarrow 0$, whereas for {\em compressible fluids} the density is variable ($\mathrm{d}\rho \neq 0$), with a finite speed of sound ($0< a < \infty$) and Mach number ($0<M<\infty$). For large flow speeds ($M > 0.1$) pressure and density are strongly coupled, especially for supersonic flows, whereas this pressure-density coupling diminishes in the incompressible flow regime ($M \rightarrow 0$), where density changes vanish ($\mathrm{d}\rho \rightarrow 0$). This acoustic degeneration as well as the pressure-velocity coupling present the major challenges for the modelling of flows with $M < 0.1$, whereas the stability in different Mach number regimes and the robust resolution of discontinuities are pertinent issues for the modelling of flows with $M>0.1$. These different mathematical characteristics and numerical requirements has made the development of algorithms that can simulate mixed compressible-incompressible two-phase fluids a difficult task.

We propose a fully-coupled pressure-based algorithm for interfacial flows at all speeds, including compressible and incompressible fluids, in which the discretised governing equations are solved for pressure, velocity and temperature. This algorithm features a conservative finite-volume discretisation of the governing equations that is identical for incompressible fluids ($\mathrm{d} \rho =0$) and compressible fluids ($\mathrm{d}\rho \neq 0$) \cite{Denner2020}. Through an appropriate linearisation, the discretised continuity equation serves as a transport equation for the density in the case of a compressible fluid, with density formulated as a function of pressure by an equation of state, and as a constraint on the velocity field for incompressible fluids, with pressure acting as a Lagrange multiplier. Hence, the proposed algorithm is applicable to incompressible, compressible and, crucially, mixed compressible-incompressible interfacial flows in all Mach number regimes. The bulk phases are represented and advected using a Volume-of-Fluid method (VOF), and the bulk phases are coupled at the interface using the acoustically-conservative interface discretisation (ACID) method \cite{Denner2018b},  which was originally proposed for fully compressible interfacial flows. Results of representative test-cases, including flows with surface tension, the pressure-driven collapse of a bubble and a shock-drop interaction, are presented to validate and scrutinise the proposed algorithm for compressible-incompressible interfacial flows.

\section{State of the art}
\label{sec:background}

As a consequence of the very different mathematical characteristics and numerical requirements of the simulation of compressible and incompressible fluids, as explained above, two different classes of algorithms have emerged: density-based algorithms and pressure-based algorithms. In the following, the main developments of both classes of algorithms related to interfacial flows are briefly reviewed in Sections \ref{sec:densitybased} and \ref{sec:pressurebased}, respectively, and, in Section \ref{sec:mixed},  currently available numerical algorithms able to simulate mixed compressible-incompressible interfacial flows are discussed.

\subsection{Density-based algorithms}
\label{sec:densitybased}
Density-based algorithms, where the sought primary variables are the density, momentum and energy of the flow, are widely applied for compressible  flows. With respect to the modelling of compressible interfacial flows, notable examples are the two-fluid Baer-Nunziato (BN) model \cite{Baer1986}, where each fluid is represented by its own set of governing equations, and the five-equation models \cite{Allaire2002, Murrone2005}, with a separate continuity equation for each phase, and shared momentum and energy equations. In these algorithms, an exact or approximate Riemann solver is employed to evaluate the fluxes at cell-faces of the computational mesh \cite{Coralic2014, Rohde2015, Garrick2017}. 
The ghost-fluid method (GFM) \cite{Fedkiw1999} has established itself as an alternative to solving a Riemann problem, partly due to its conceptual simplicity. Recent developments have also seen a combination of GFM with Riemann solvers \cite{Liu2003, Liu2017} to improve the stability of simulations with strong shock-interface interactions and compressible gas-liquid flows. 

Since pressure is not directly solved for with a density-based algorithm, the pressure field has to be reconstructed based on the applied thermodynamic model via an equation of state, which may be difficult in interfacial cells where two bulk phases coexist \cite{Allaire2002, Murrone2005, Abgrall2001}, and which requires hydrodynamically and thermodynamically plausible fluid properties.
While density-based algorithms are naturally suited for compressible flows, they are poorly suited for low-Mach number flows \cite{Karimian1994, Wesseling2001}, where the coupling of pressure and density vanishes. Although density-based algorithms have been applied to low-Mach number flows with some success, this requires computationally expensive pre-conditioning techniques \cite{Turkel1993}. 

A different approach for single-phase flows at all speeds was pursued by van der Heul et al.~\cite{vanderHeul2003}, in which the energy equation is reformulated as an equation for pressure, while the continuity equation still serves as an equation for density. Similar methods were subsequently presented by Park and Munz \cite{Park2005} and Cordier et al.~\cite{Cordier2012}. These algorithms formally converge to the incompressible flow equations in the limit of zero Mach number ($M \rightarrow 0$) \cite{vanderHeul2003, Park2005}, with the reformulated energy equation enforcing a divergence-free velocity field. Boger et al.~\cite{Boger2014} extended the work of Park and Munz \cite{Park2005} to interfacial flows. 

\subsection{Pressure-based algorithms}
\label{sec:pressurebased}
Pressure-based algorithms, in which the continuity equation is cast into a discretised equation for pressure (with pressure acting as the Lagrange multiplier that enforces $\mathbf{\nabla} \cdot \mathbf{u} \rightarrow 0$ for $M \rightarrow 0$), while density is either constant (incompressible fluids) or evaluated using a suitable equation of state (compressible fluids), are preferably applied to simulate the flow of incompressible fluids and weakly compressible flows, and may yield significant advantages for low-Mach number compressible flows, since the acoustic degeneration at low Mach numbers does not pose a problem \cite{Hauke1998}. The success of pressure-based algorithms is facilitated by the unique role of pressure in all Mach number regimes \cite{Hauke1998}, with the pressure-velocity coupling dominant at low Mach numbers and the pressure-density coupling dominant at high Mach numbers \cite{VanDoormaal1987, Karki1989}, and the convenient fact that the fully-conservative formulation of the governing conservation laws can still be satisfied accurately \cite{Denner2020, VanDoormaal1987}.

The majority of pressure-based algorithms for incompressible interfacial flows are founded on pressure-correction methods, such as projection methods \cite{Chorin1967, Bell1989}, the SIMPLE method \cite{Patankar1980} and its subsequent derivatives, or the PISO method \cite{Issa1985}, or fully-coupled algorithms \cite{Denner2014a}.
Pressure-based algorithms for compressible flows are much less prominent in the literature than their density-based counterparts, partly because deriving stable and efficient numerical schemes for the transonic regime \cite{Denner2018c} as well as formulating consistent shock-capturing schemes have proven difficult for pressure-based algorithms \cite{Wesseling2001, Bijl1998}. Nonetheless, because pressure plays an important role in all Mach number regimes, pressure-based algorithms have a distinct potential for applications in all Mach number regimes or in which the Mach number varies strongly. Starting with the work of Harlow and Amsden \cite{Harlow1971a}, a variety of pressure-based algorithms for compressible single-phase flows has been developed \cite{Karimian1994, VanDoormaal1987, Denner2018c, Darwish2014, Xiao2017}. Denner et al.~\cite{Denner2020} presented a fully-coupled pressure-based algorithm able to predict single-phase flows of incompressible, ideal-gas and real-gas fluids at all speeds ($0\leq M \leq 239$) accurately and robustly, using the same fully-coupled pressure-based algorithm and conservative finite-volume discretisation. 

Despite these developments for single-phase flows, it was only recently that the first pressure-based algorithm in conservative form for compressible interfacial flows at all speeds was proposed \cite{Denner2018b}. This algorithm was proposed in conjunction with a new interface discretisation method that retains the acoustic features of the flow, without the need to employ a Riemann solver, and has been shown to be a promising alternative to traditional density-based algorithms, with the pressure-based formulation of the governing equations facilitating the definition of consistent mixture rules at the interface that apply naturally to flows in all Mach number regimes. This algorithm was subsequently extended to polytropic interfacial flows \cite{Denner2020a}, {\em i.e.}~treating the flow as a polytropic process.

\subsection{Algorithms for compressible-incompressible flows}
\label{sec:mixed}
Caiden et al.~\cite{Caiden2001} were the first to propose a numerical algorithm dedicated to the simulation of general compressible-incompressible flows, solving different governing equations for the compressible fluid and the incompressible fluid, coupling the bulk phases at the interface using the GFM. A similar method to describe the behaviour of compressible bubbles in an incompressible fluid was subsequently presented by Aanjaneya et al.~\cite{Aanjaneya2013}. Wadhwa et al.~\cite{Wadhwa2005} proposed a method for computing incompressible liquid drops in a compressible gas, representing the interface as a matching moving mesh and solving the flow of the incompressible fluid using the artificial compressibility method of Chorin \cite{Chorin1967}. Others proposed algorithms for compressible-incompressible flows applying the same governing equations in non-conservative form in both fluids \cite{Billaud2011, Caltagirone2011, Yamamoto2018} and these algorithms are, in general, only applicable to incompressible and low Mach number flows. However, the application of the governing equations in conservative form together with a conservative discretisation is a prerequisite for an accurate prediction of flows in all Mach number regimes \cite{Hou1994}, in particular shock waves, rarefaction fans and contact discontinuities. The density-based algorithm of Boger et al.~\cite{Boger2014} is readily applicable to interfacial flows in all Mach number regimes, including the incompressible limit ($M \rightarrow 0$), yet incompressible fluids ($M=0$) have not been considered in the context of such an algorithm. 

In addition to the difficulties encountered when developing numerical methods for interfacial flows, such as a precise discrete force balance between surface tension and the pressure gradient, Caiden et al.~\cite{Caiden2001} and Billaud et al.~\cite{Billaud2011} identified the transmission of waves at the interface, as well as physically realistic and compatible discrete formulations of the governing equations describing the incompressible fluid and the compressible fluid, as the main additional difficulties associated with an accurate prediction of compressible-incompressible interfacial flows. The presented algorithm addresses and resolves these difficulties, as demonstrated by the chosen validation cases.

\section{Governing equations}
\label{sec:governingEq}

The conservation laws governing both incompressible and compressible fluid flow at all speeds are the conservation of mass, momentum and energy, given as
\begin{align}
\frac{\partial \rho}{\partial t} + \mathbf{\nabla} \cdot (\rho
\mathbf{u}) &= 0
\label{eq:continuity} \\
\frac{\partial (\rho \mathbf{u})}{\partial t} + \mathbf{\nabla} \cdot
(\rho \mathbf{u} \otimes \mathbf{u}) &= - \mathbf{\nabla} p +
\mathbf{\nabla} \cdot \boldsymbol{\tau} + \mathbf{S} 
\label{eq:momentum} \\
\frac{\partial (\rho h)}{\partial t} + \mathbf{\nabla} \cdot \left( \rho
\mathbf{u} h \right) &= \frac{\partial p}{\partial t} - \mathbf{\nabla}
\cdot \mathbf{q} + \mathbf{\nabla} \cdot \left(\boldsymbol{\tau} \cdot
\mathbf{u} \right) + \mathbf{S} \cdot \mathbf{u},
\label{eq:energy}
\end{align}
respectively, where $t$ is time, $\mathbf{u}$ is the velocity vector, $p$ is pressure, $\rho$ is the density and $h$ is the specific total enthalpy. The stress tensor $\boldsymbol{\tau}$ for the considered Newtonian fluids is given as
\begin{equation}
\boldsymbol{\tau} = \mu \left( \mathbf{\nabla} \mathbf{u} + (\mathbf{\nabla} \mathbf{u})^T \right) - \frac{2}{3} \mu \left(\mathbf{\nabla} \cdot \mathbf{u} \right) \mathbf{I},
\end{equation}
where $\mu$ is the dynamic viscosity. The heat flux due to thermal conduction is typically described by Fourier's law as $\mathbf{q} = - k \mathbf{\nabla}T$, with $k$ the thermal conductivity and $T$ the temperature. All external forces applied to the flow, {\em e.g.}~the force due to surface tension, are gathered in the volumetric source term $\mathbf{S}$.

The Volume-of-Fluid (VOF) method \cite{Hirt1981} is adopted to capture the fluid interface between two immiscible bulk phases, applying an indicator function field $\zeta$, defined as
\begin{equation}
\zeta (\mathbf{x}) =
\begin{cases}
0 & \text{if} \ \mathbf{x} \in \Omega_\mathrm{a} \\
1 & \text{if} \ \mathbf{x} \in \Omega_\mathrm{b},
\end{cases} \label{eq:colourDef}
\end{equation}
where $\Omega_\mathrm{a}$ and $\Omega_\mathrm{b}$ are the subdomains occupied by fluid a and fluid b, respectively, and $\Omega = \Omega_\mathrm{a} \cup \Omega_\mathrm{b}$ is the computational domain. Because the interface is a material front propagating with the flow, the indicator function $\zeta$ is advected with the underlying fluid velocity by the advection equation
\begin{equation}
\frac{\partial \zeta}{\partial t} + \mathbf{u} \cdot \mathbf{\nabla} \zeta  = 0 . \label{eq:vofAdvection}
\end{equation}

\section{Thermodynamic closure}
\label{sec:thermodynamicClosure}

The governing conservation laws given above require closure by an appropriate thermodynamic model, defining the thermodynamic properties of the fluids.  Following the approach recently proposed by Denner et al.~\cite{Denner2020}, the density $\rho$ and the specific isobaric heat capacity $c_p$ are defined by a set of input parameters ($\rho_0$, $c_{p,0}$, $c_{v,0}$ and $\Pi_0$) that enables the formulation of a unified thermodynamic closure for incompressible and compressible fluids.

For a compressible fluid, the stiffened-gas model \cite{Harlow1971,LeMetayer2004} is applied, in which the density is defined as
\begin{equation}
\rho =  \frac{p + \gamma_0 \Pi_0}{R_0 \, T} , \label{eq:rhoComp}
\end{equation}
where $\Pi_0$ is a material-dependent pressure constant, $R_0 = c_{p,0}-c_{v,0}$ is the specific gas constant, with the constant reference specific isobaric heat capacity $c_{p,0}$ and the constant reference specific isochoric heat capacity $c_{v,0}$, and $\gamma_0 = c_{p,0}/c_{v,0}$ is the heat capacity ratio. The specific isobaric heat capacity is given as \cite{Denner2018b}
\begin{equation}
c_p = c_{p,0} \frac{p + \Pi_0}{p + \gamma_0 \Pi_0} , \label{eq:cpComp}
\end{equation} 
and the specific total enthalpy is defined as $h = c_p \, T + \mathbf{u}^2/2$.
The speed of sound follows as
\begin{equation}
a = \sqrt{\gamma_0 \, \frac{p + \Pi_0}{\rho}} .
\label{eq:soundSpeed}
\end{equation}
For $\Pi_0=0$, the stiffened-gas model describes a calorically perfect ideal gas, with $\rho = p/(R_0 \, T)$ and $c_p = c_{p,0}$.

The density of an incompressible fluid is, by definition, constant and given as
\begin{equation}
\rho = \rho_{0} , \label{eq:rhoIncomp}
\end{equation}
with $\rho_{0}$ a predefined density value. The specific isobaric heat capacity $c_p$ of an incompressible fluid is also assumed to be constant, with
\begin{equation}
c_p = c_{p,0} . \label{eq:cpIncomp}
\end{equation}

In order to incorporate compressible and incompressible fluids in the same numerical framework, the definitions for the density $\rho$ and the specific isobaric heat capacity $c_p$ are unified by the binary operator  $\mathcal{C}$, given as
\begin{equation}
\mathcal{C} = 
\begin{cases}
1 \, ,  \ \text{for compressible fluids}\\
0 \, , \ \text{for incompressible fluids}.
\end{cases} \label{eq:operatorC}
\end{equation}
This binary operator is used as a coefficient for the compressible part and, analogously, $1-\mathcal{C}$ is used as a coefficient for the incompressible part of the unified closure model. The density for a given fluid is then defined, based on Eqs.~(\ref{eq:rhoComp}) and (\ref{eq:rhoIncomp}), as
\begin{equation}
\rho = \mathcal{C} \, \frac{p + \gamma_0 \Pi_0}{R_0 \, T} + (1-\mathcal{C})
\,  \rho_{0} ,
\label{eq:rhoFull}
\end{equation}
and the specific isobaric heat capacity is defined, based on Eqs.~(\ref{eq:cpComp}) and (\ref{eq:cpIncomp}), as
\begin{equation}
c_p = c_{p,0} \left[\mathcal{C} \, \frac{p + \Pi_0}{p + \gamma_0 \Pi_0}  +
(1-\mathcal{C}) \right] . \label{eq:cpFull}
\end{equation}
The speed of sound is given, following Eq.~(\ref{eq:soundSpeed}), as
\begin{equation}
a = \mathcal{C} \, \sqrt{\gamma_0 \frac{p+\Pi_0}{\rho}} + (1-\mathcal{C}) \,
a_\infty,
\end{equation}
where $a_\infty$ is a very large velocity (here: $a_\infty=10^{32} \, \textup{m}\, \textup{s}^{-1}$) that represents the infinite speed of sound of incompressible fluids and ensures a computationally meaningful definition of the speed of sound throughout the computational domain. The appropriate formulations of the density $\rho$, the specific isobaric heat capacity $c_p$ and the speed of sound $a$ for incompressible fluids ($\mathcal{C}=0$), perfect-gas fluids ($\mathcal{C}=1$, $\Pi_0 = 0$) and stiffened-gas fluids ($\mathcal{C}=1$, $\Pi_0 > 0$) are readily recovered.

\section{Numerical framework}
\label{sec:numericalFramework}
The proposed numerical algorithm is based on a fully-coupled pressure-based algorithm for incompressible and compressible fluids with a finite-volume discretisation of the governing equations \cite{Denner2020}. The numerical framework is predicated on a collocated variable arrangement, meaning that the primary
solution variables $p$, $\mathbf{u}$ and $T$, as well as all fluid properties, are stored at the centre of the mesh cells. First, the finite-volume discretisation underpinning the numerical algorithm is briefly discussed (Section \ref{sec:fvm}), followed by the definition of the advecting velocity (Section \ref{sec:mwi}) and the discretisation of the governing equations (Section \ref{sec:disc}). The particular form the discretised governing equations assume in the specific case of the incompressible limit is discussed in Section \ref{sec:incompLimit} and the solution procedure is described in Section \ref{sec:solution}.

\subsection{Finite-volume discretisation}
\label{sec:fvm}
Considering, for example, the advective-diffusive transport of a general fluid variable, $\phi$, is given as
\begin{equation}
\frac{\partial (\rho \phi)}{\partial t} + \mathbf{\nabla} \cdot (\rho \mathbf{u} \phi) = \mathbf{\nabla} \cdot (\Gamma_\phi  \, \mathbf{\nabla} \phi) , \label{eq:transportEq}
\end{equation}
where $t$ is the time, $\rho$ is the fluid density, $\mathbf{u}$ is the fluid velocity and $\Gamma_\phi$ is the diffusion coefficient of $\phi$. 
Reformulating Eq.~(\ref{eq:transportEq}) in its integral form with respect to the control volume $V$, given as
\begin{equation}
\iiint_V \frac{\partial (\rho \phi)}{\partial t} \, \textup{d}V + \iiint_V
 \mathbf{\nabla} \cdot (\rho \mathbf{u} \phi) \, \textup{d}V = \iiint_V 
\mathbf{\nabla} \cdot (\Gamma_\phi  \, \mathbf{\nabla} \phi)  \, \textup{d}V,
\label{eq:transportEqInt}
\end{equation}
allows a rather straightforward finite-volume discretisation of the advective-diffusive transport of $\phi$. In the interest of simplicity and brevity, the mesh is henceforth assumed to be  Cartesian with a local mesh spacing $\Delta x$. Extending the presented methods to unstructured meshes is easily achieved by introducing  corrections for mesh skewness and non-orthogonality, as for instance described in \cite{Denner2020}.

The transient term is discretised in the following using the Second-Order Backward Euler scheme, which is given for cell $P$ as
\begin{equation}
\iiint_{V} \frac{\partial \Phi}{\partial t} \, \textup{d}V \approx \frac{3
\Phi_P^{(n+1)} - 4 \Phi_P^{(t-\Delta t)} + \Phi_P^{(t-2\Delta t)}}{2 \Delta t} \,
V_P + \mathcal{O}(\Delta t^2),
\end{equation} 
where $\Phi = \rho \phi$, $\Delta t$ is the time-step, superscript $(n+1)$ denotes the implicitly sought solution, superscript $(t-\Delta t)$ denotes the previous time-level and superscript $(t-2\Delta t)$ denotes the previous previous time-level. 

\begin{figure}[t]
\begin{center}
\includegraphics[scale=0.9]{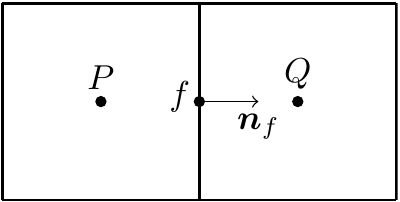}
\label{fig:discGeoGeneral}
\caption{Schematic illustration of mesh cell $P$ with its neighbour mesh cell $Q$ and the shared mesh face $f$, where $\mathbf{n}_f$ is the unit normal vector of face $f$, pointing out of cell $P$.}
\label{fig:discGeo}
\end{center}
\end{figure}

The discrete form of the advection term is obtained by applying the divergence theorem,
\begin{align}
\iiint_V \mathbf{\nabla} \cdot (\rho \mathbf{u} \phi) \, \textup{d}V & =
\oiint_{\partial V} \rho \mathbf{u} \phi \, \textup{d}\mathbf{\Sigma} , 
\end{align}
where $\mathbf{\Sigma}$ is the outward-pointing surface vector on the surface $\partial V$ of control volume $V$. Since the surface of a discrete control volume is constituted by a finite number of flat faces $f$ of area $A_f$, as illustrated in Fig.~\ref{fig:discGeo}, and applying the midpoint rule \cite{Ferziger2020, Moukalled2016}, the advection term follows as 
\begin{align}
\oiint_{\partial V} \rho \mathbf{u} \phi \, \textup{d}\mathbf{\Sigma}  & \approx \sum_f \rho_f
\vartheta_f \phi_f A_f = \sum_f \dot{m}_f \phi_f .
\end{align}
The face value $\phi_f$ is typically not readily available and has to be interpolated from the cell-centred values, which in a finite-volume sense represent the values of $\phi$ averaged over the respective cell volume, using an appropriate interpolation scheme.
The advecting velocity $\vartheta_f = \mathbf{u}_f \cdot \mathbf{n}_f$, where $\mathbf{n}_f$ is the normal vector of face $f$ (pointing outwards with respect to cell $P$, see Fig.~\ref{fig:discGeo}), represents the velocity normal to face $f$ and forms, together with the face density $\rho_f$ and the face area $A_f$, the mass flux $\dot{m}_f = \rho_f \vartheta_f A_f$ through face $f$. In the context of the presented algorithm, the advecting velocity $\vartheta_f$ is determined using a momentum-weighted interpolation from the cell-centred values of velocity and pressure, as further discussed in Section \ref{sec:mwi}. Since the values of density and velocity reside at the cell centres, a suitable interpolation must be applied, such as upwind differencing, central differencing or a TVD scheme.

Similar to the advection term, applying the divergence theorem in conjunction with the midpoint rule, the diffusion term of the transport equation (\ref{eq:transportEqInt}) is discretised as
\begin{equation}
\iiint_V  \mathbf{\nabla} \cdot (\Gamma_\phi  \, \mathbf{\nabla} \phi)  \, \textup{d}V 
\approx \sum_f \Gamma_{\phi,f} \left( \mathbf{\nabla} \phi|_f \cdot \mathbf{n}_{f} \right) A_f,
\end{equation}
where the diffusion coefficient $\Gamma_\phi$ at face $f$ is defined by a harmonic average of the cell-centred values as \cite{Ferziger2003}
\begin{equation}
\Gamma_{\phi,f} = \frac{2}{\Gamma_{\phi,P}^{-1} + \Gamma_{\phi,Q}^{-1}} , \label{eq:diffCoeffFace}
\end{equation}
with cell $Q$ the neighbour cell of cell $P$, as illustrated in Fig.~\ref{fig:discGeo}.
The face-centred gradient of $\phi$ along the normal vector $\mathbf{n}_f$ is approximated with second-order accuracy as
\begin{equation}
\mathbf{\nabla} \phi|_f \cdot \mathbf{n}_{f} \approx
\frac{\phi_Q-\phi_P}{\Delta x},
\label{eq:gradFace}
\end{equation}
where $\Delta x$ is the mesh spacing.

Applying the divergence theorem, the spatial gradient of $\phi$ averaged over the cell volume $V_P$ is readily computed as
\begin{equation}
\mathbf{\nabla} \phi_P \approx \frac{1}{V_P} \sum_f \phi_f \, \mathbf{n}_f \, A_f.
\label{eq:gradGauss}
\end{equation}

\subsection{Advecting velocity}
\label{sec:mwi}
The momentum-weighted interpolation (MWI) is applied to define an {\em advecting velocity} $\vartheta=\mathbf{u}_f \cdot \mathbf{n}_f$ at cell faces, which is used in the discretised advection terms of the governing equations. MWI provides a robust pressure-velocity coupling for flows with low Mach numbers and flows of incompressible fluids by introducing a cell-to-cell pressure coupling and by applying a low-pass filter acting on the third derivative of pressure \cite{Ferziger2002, Bartholomew2018}, thus avoiding pressure-velocity decoupling due to the collocated variable arrangement.  

Following the unified formulation of the MWI proposed by Bartholomew et al.~\cite{Bartholomew2018}, the definition of the advecting velocity includes modifications to the original formulation of the MWI, as introduced by Rhie and Chow \cite{Rhie1983}, to account for large density ratios and source terms occurring in interfacial flows, and the transient nature of the considered problems. As demonstrated by Bartholomew et al.~\cite{Bartholomew2018}, only the {\em driving} pressure gradient $\mathbf{\nabla}P = \mathbf{\nabla} p - \mathbf{S}$, which is the pressure gradient associated with the flow field, should be coupled to the velocity field, whereas source terms and other external contributions should not be coupled to the velocity field. Taking this into account, the advecting velocity $\vartheta_f$ at face $f$ is given as
\begin{equation}
\begin{split}
\vartheta_f = \overline{\mathbf{u}}_{f} \cdot \mathbf{n}_{f} &- \hat{{d}}_f
\left[ \mathbf{\nabla} P_f \cdot \mathbf{n}_{f} - \frac{\rho_f^\ast}{2} \, \left(\frac{\mathbf{\nabla} P_P}{\rho_P} + \frac{\mathbf{\nabla} P_Q}{\rho_Q} \right) \cdot \mathbf{n}_{f} \right]  \\ &+  \hat{{d}}_f \, \frac{\rho^{\ast (t-\Delta t)}_f}{\Delta t} \left(\vartheta^{(t-\Delta t)}_f - \overline{\mathbf{u}}_{f}^{(t-\Delta t)} \cdot \mathbf{n}_{f} \right)  ,
\label{eq:advVel}
\end{split}
\end{equation} 
where the face velocities $\overline{\mathbf{u}}_f$ and $\overline{\mathbf{u}}_f^{(t-\Delta t)}$ are obtained by linear interpolation, and
\begin{equation}
\mathbf{\nabla} P_f \cdot \mathbf{n}_{f} \approx \frac{P_Q-P_P}{\Delta x}.
\end{equation}
The face density $\rho_f^\ast$ is interpolated by a harmonic average, which is necessary for a consistent definition of the coefficient of the pressure term as well as the efficacy of the density weighting \cite{Bartholomew2018}. The coefficient $\hat{{d}}_f$ is defined as \cite{Denner2014a, Bartholomew2018}
\begin{equation}
\hat{{d}}_f = \dfrac{\left(\dfrac{V_P}{{e}_P} +
\dfrac{V_Q}{{e}_Q} \right)}{2+\dfrac{\rho^\ast_f}{\Delta t} \,
\left(\dfrac{V_P}{{e}_P} +
\dfrac{V_Q}{{e}_Q} \right)} ,
\end{equation}
where ${e}_P$ and ${e}_Q$ are the sum of the coefficients of the primary variable $\mathbf{u}$ arising from the advection and shear stress terms of the momentum equations \cite{Denner2020}. 

The MWI formulation given in Eq.~(\ref{eq:advVel}) is independent of the applied time-step and the error in kinetic energy introduced by the MWI is proportional to $\Delta x^3$ \cite{Bartholomew2018}. Hence, the convergence of the second-order accurate finite-volume method remains unaffected.  
The density weighting applied to the cell-centred pressure gradient in Eq.~(\ref{eq:advVel}) has been shown to yield robust results for flows with large and abrupt changes in density \cite{Bartholomew2018}, demonstrated for incompressible interfacial flows with a density ratio of up to $10^{24}$ \cite{Denner2014a, Denner2015}. Including the transient term has previously been shown to be important for a correct temporal evolution of pressure waves \cite{Xiao2017, Bartholomew2018}, which is particularly pertinent for acoustic effects in compressible flows.

\subsection{Discretised governing conservation laws} 
\label{sec:disc}
Since the discretisation of the governing conservation laws is identical for single-phase and interfacial flows, the following presentation of the discretisation of the governing conservation laws focuses on single-phase flows. The extension of this discretisation to interfacial flows by an appropriate definition of the fluid properties and using the ACID method is described in Section \ref{sec:interface}.

Applying the numerical schemes described in the previous sections, the discretised continuity equation (\ref{eq:continuity}) for cell $P$ is given as
\begin{equation}
\left. \frac{\partial \rho}{\partial t} \right|_P V_P + \sum_f (\rho_f \, \vartheta_f)^{(n+1)} \, A_f = 0 ,
\label{eq:continuityDisc}
\end{equation}
where superscript $(n+1)$ denotes implicitly sought solutions.
A Newton linearisation \cite{Denner2018c, Xiao2017} is applied to the advection term of the discretised continuity equation, given as
\begin{equation}
(\rho_f \, \vartheta_f)^{(n+1)} \approx \rho^{(n)}_f \, \vartheta_f^{(n+1)} +
\rho^{(n+1)}_f \, \vartheta_f^{(n)} - \rho^{(n)}_f \,
\vartheta_f^{(n)} ,
\label{eq:newtonLinConti}
\end{equation}
where superscript $(n)$ denotes the latest available solution. The Newton linearisation facilitates a smooth transition from low to high Mach number regions \cite{VanDoormaal1987, Denner2018c} and provides an implicit pressure-velocity coupling for flows at low Mach numbers and flows of incompressible fluids \cite{Denner2020, Xiao2017}.

The discretised momentum equations (\ref{eq:momentum}) of cell $P$ are given as
\begin{equation}
\begin{split}
& \left. \frac{\partial \rho u_j}{\partial t} \right|_P V_P + \sum_f
(\rho_f \, \vartheta_f \, u_{j,f})^{(n+1)} \, A_f  =  - \sum_f \overline{p}_f^{(n+1)} \, {n}_{j,f} \, A_f \\ & + \sum_f \mu_f^\ast \left[ \frac{u_{j,Q}^{(n+1)} - u_{j,P}^{(n+1)}}{\Delta x} + \left. \overline{\frac{\partial u_i}{\partial x_j}} \right|_f^{(n)} \, {n}_{i,f}  - \frac{2}{3}  \left. \overline{\frac{\partial u_k}{\partial  x_k}} \right|_f^{(n)} \, {n}_{i,f}  \right] A_f + S_{j,P} \, V_P  ,
\end{split} \label{eq:momentumDisc}
\end{equation}
where $\overline{\square}_f$ denotes linear interpolation of the values at the adjacent cell centres and
\begin{equation}
\begin{split}
(\rho_f \, \vartheta_f \, u_{j,f})^{(n+1)} &\approx \rho^{(n)}_f \, \vartheta_f^{(n)} \, u_{j,f}^{(n+1)} + \rho^{(n)}_f \, \vartheta_f^{(n+1)} \, u_{j,f}^{(n)} \\ &+ \rho^{(n+1)}_f \, \vartheta_f^{(n)} \, u_{j,f}^{(n)} - 2 \, \rho^{(n)}_f \,
\vartheta_f^{(n)} \, u_{j,f}^{(n)}.
\end{split}
\label{eq:linearisationVelocityAdv}
\end{equation}
The discretised energy equation (\ref{eq:energy}) of cell $P$ is given as
\begin{equation}
\begin{split} 
& \left. \frac{\partial \rho h}{\partial t} \right|_P V_P  + \sum_f (\rho_f \, \vartheta_f \, h_{f})^{(n+1)} \, A_f = \left. \frac{\partial p}{\partial t} \right|_P V_P + \sum_f k_f^\ast \frac{T_Q^{(n+1)}-T_P^{(n+1)}}{\Delta x}  \, A_f \\ & +  \sum_f \overline{u}_{i,f}^{(n+1)} \, \mu_f^\ast \left[ \left(\left. \overline{\frac{\partial u_j}{\partial x_i}} \right|_f^{(n)} + \left. \overline{\frac{\partial u_i}{\partial x_j}} \right|_f^{(n)} \right) -  \frac{2}{3} \left. \overline{\frac{\partial u_k}{\partial x_k}} \right|_f^{(n)}  \right] {n}_{j,f} \, A_f \\ &+ \, u_{i,P}^{(n+1)} \, S_{i,P} \, V_P ,
\end{split} \label{eq:energyDisc}
\end{equation}
with
\begin{equation}
\begin{split}
(\rho_f \, \vartheta_f \, h_f)^{(n+1)} &\approx \rho^{(n)}_f \, \vartheta_f^{(n)} \, h_f^{(n+1)} + \rho^{(n)}_f \, \vartheta_f^{(n+1)} \, h_f^{(n)} \\ &+ \rho^{(n+1)}_f \, \vartheta_f^{(n)} \, h_f^{(n)} - 2 \, \rho^{(n)}_f \,
\vartheta_f^{(n)} \, h_f^{(n)},
\end{split}
\label{eq:linearisationEnthalpyAdv}
\end{equation}
and 
\begin{equation}
h^{(n+1)} = c_p \, T^{(n+1)} + \frac{\mathbf{u}^{(n),2}}{2}.
\label{eq:enthalpyImplemented}
\end{equation}
The advecting velocity $\vartheta_f$ is given by Eq.~(\ref{eq:advVel}) and is the same in all equations, ensuring a consistent transport of the conserved quantities. Following the work of Ferziger \cite{Ferziger2003}, the viscosity $\mu_f^\ast$ and the thermal conductivity $k_f^\ast$ at face $f$ are given by a harmonic average of the values at the adjacent cell centres.

Previous studies \cite{Denner2018b,Denner2018c} have demonstrated substantial improvements with respect to the performance and stability of the numerical solution algorithm associated with a Newton linearisation of all governing equations. In particular, such a linearisation enables a smooth transition from low to high Mach number regions \cite{Karimian1994,Denner2018c,Kunz1999}. To this end, the advection terms of the governing equations as well as the transient terms of the momentum and energy equations are discretised using a Newton linearisation, see Eqs.~(\ref{eq:linearisationVelocityAdv}) and (\ref{eq:linearisationEnthalpyAdv}),  as described in more detail in \cite{Denner2018c}.

\subsection{Incompressible limit}
\label{sec:incompLimit}

For flows in the incompressible limit, with $M \rightarrow 0$, the density is constant along the fluid particle trajectories \cite{Chorin1993}, with
\begin{equation}
\frac{\mathrm{D}\rho}{\mathrm{D}t} = \frac{\partial \rho}{\partial t} + \mathbf{u} \cdot \mathbf{\nabla} \rho = 0.
\label{eq:materialDerivativeDensity}
\end{equation}
Consequently, the continuity equation is no longer effective as a transport equation for density, but becomes a constraint on the velocity field with $\mathbf{\nabla} \cdot \mathbf{u} \rightarrow 0$ \cite{Chorin1993}, enforced by pressure which acts as a Lagrange multiplier \cite{Toutant2017}.
Inserting Eq.~(\ref{eq:materialDerivativeDensity}) into
Eqs.~(\ref{eq:continuity})--(\ref{eq:energy}), the governing conservation laws for flows in the incompressible limit reduce to
\begin{align}
\mathbf{\nabla} \cdot  \mathbf{u} &= 0
\label{eq:continuityImp} \\
\rho  \left( \frac{\partial \mathbf{u}}{\partial t} + \mathbf{\nabla} \cdot
(\mathbf{u} \otimes \mathbf{u})\right) &= - \mathbf{\nabla} p +
\mathbf{\nabla} \cdot \boldsymbol{\tau} + \mathbf{S} 
\label{eq:momentumImp} \\
\rho  \left(\frac{\partial h}{\partial t} + \mathbf{\nabla} \cdot \left(
\mathbf{u} h \right)\right) &= \frac{\partial p}{\partial t} - \mathbf{\nabla}
\cdot \mathbf{q} + \mathbf{\nabla} \cdot \left(\boldsymbol{\tau} \cdot
\mathbf{u} \right) + \mathbf{S} \cdot \mathbf{u}.
\label{eq:energyImp}
\end{align}

Inserting $\rho=\rho_0$, see Eq.~(\ref{eq:rhoIncomp}), into the discretised governing conservation laws presented above, the discretised continuity equation (\ref{eq:continuityDisc}) reduces to
\begin{equation}
\sum_f \vartheta_f \,  A_f = 0 \ ,
\label{eq:continuityDiscImp}
\end{equation}
the discretised momentum equations (\ref{eq:momentumDisc}) become
\begin{equation}
\begin{split}
& \rho_P \left(\left. \frac{\partial u_j}{\partial t} \right|_P V_P + \sum_f
\vartheta_f \, u_{j,f} \, A_f\right) =  - \sum_f \overline{p}_f \, {n}_{j,f} \, A_f \\ & + \sum_f \mu_f \left[\left(\left.
{\frac{\partial u_j}{\partial x_i}} \right|_f +
\left. {\frac{\partial u_i}{\partial x_j}} \right|_f \right) \,
 - \frac{2}{3}  \left. \frac{\partial u_k}{\partial
 x_k} \right|_f \right] {n}_{i,f} \, A_f + S_{j,P} \, V_P , 
\end{split} \label{eq:momentumDiscImp}
\end{equation}
and the discretised energy equation (\ref{eq:energyDisc}) becomes
\begin{equation}
\begin{split} 
& \rho_P \left(\left. \frac{\partial h}{\partial t} \right|_P V_P  + \sum_f  \vartheta_f \, h_{f} \, A_f \right) = \left. \frac{\partial p}{\partial t} \right|_P V_P + \sum_f k_f \left. \frac{\partial T}{\partial x_i} \right|_f  {n}_{i,f} \, A_f \\ & +  \sum_f \overline{u}_{i,f} \mu_f \left[ \left(\left. {\frac{\partial u_j}{\partial x_i}} \right|_f + \left. {\frac{\partial u_i}{\partial x_j}} \right|_f \right) -  \frac{2}{3} \left. \frac{\partial u_k}{\partial x_k} \right|_f  \right] {n}_{j,f} \, A_f + \, u_{i,P} \, S_{i,P} \, V_P .
\end{split} \label{eq:energyDiscImp}
\end{equation}
These are precisely the governing conservation laws of the incompressible limit, Eqs.~(\ref{eq:continuityImp})--(\ref{eq:energyImp}), discretised with the schemes as discussed above. In addition, Eqs.~(\ref{eq:continuityDiscImp}) and (\ref{eq:momentumDiscImp}) are identical to the discretised continuity and momentum equations of the fully-coupled pressure-based algorithm for incompressible interfacial flows of Denner et al.~\cite{Denner2014a}.

\subsection{Solution procedure}
\label{sec:solution}
The discretised governing conservation laws are solved simultaneously in a single linear system of equations, $\mathbf{A} \cdot \boldsymbol{\phi} = \mathbf{b}$. Following Denner et al.~\cite{Denner2020}, the system of equations is solved for the primary solution variables $\chi$, which are the pressure $p$, the velocity vector $\mathbf{u}$ and the temperature $T$. For a three-dimensional computational mesh with $N$ cells, the linear system of governing equations is given as
\begin{equation}
\begin{pmatrix}
\mathbf{A}_p & \mathbf{A}_u & \mathbf{A}_v & \mathbf{A}_w & \mathbf{0} \\
\mathbf{B}_p & \mathbf{B}_u & \mathbf{B}_v & \mathbf{B}_w & \mathbf{0} \\
\mathbf{C}_p & \mathbf{C}_u & \mathbf{C}_v & \mathbf{C}_w & \mathbf{0} \\
\mathbf{D}_p & \mathbf{D}_u & \mathbf{D}_v & \mathbf{D}_w & \mathbf{0} \\
\mathbf{E}_p & \mathbf{E}_u & \mathbf{E}_v & \mathbf{E}_w & \mathbf{E}_T \\
\end{pmatrix} \cdot
\begin{pmatrix}
\boldsymbol{\phi}_p \\
\boldsymbol{\phi}_u \\
\boldsymbol{\phi}_v \\
\boldsymbol{\phi}_w \\ 
\boldsymbol{\phi}_T
\end{pmatrix} = \mathbf{b} \ ,
\label{eq:eqsys}
\end{equation} 
where the coefficient submatrix $\mathbf{A}_\chi$ of size $N \times N$ holds the coefficients of the primary variable $\chi$ associated with the continuity equation (\ref{eq:continuityDisc}), submatrices $\mathbf{B}_\chi$, $\mathbf{C}_\chi$, $\mathbf{D}_\chi$, all of size $N \times N$, hold the coefficients of the primary variable $\chi$ associated with the momentum equations (\ref{eq:momentumDisc}), for the $x$-, $y$- and $z$-axes of the Cartesian coordinate system, and submatrix $\mathbf{E}_\chi$, also of size $N \times N$, contains the coefficients of the primary variable $\chi$ associated with the energy equation (\ref{eq:energyDisc}). The subvector $\boldsymbol{\phi}_\chi$ of length $N$ holds the solution of the primary variable $\chi$. All contributions from previous non-linear iterations or previous time-levels are gathered in the right-hand side vector $\mathbf{b}$ of length $5N$. 

The solution procedure performs nonlinear iterations in which the linear system of governing equations (\ref{eq:eqsys}) is solved using the {\em Block-Jacobi} preconditioner and the {\em BiCGSTAB} solver of the software library PETSc  \cite{petsc-user-ref}, as described in detail by Denner et al.~\cite{Denner2018b}. No underrelaxation of the governing equations is required.

\section{Interface treatment}
\label{sec:interface}
The discretised governing equations are extended to interfacial flows using an interface advection method (Section \ref{sec:interfaceAdvection}), an appropriate definition of the fluid properties in the interface region (Section \ref{sec:fluidProp}) and the acoustically-conservative interface discretisation (ACID)\footnote{The term ``acoustically-conservative'' refers to the acoustic properties of this discretisation method in the context of fully compressible flows and is not indicative of its application to incompressible fluids.} \cite{Denner2018b} (Section \ref{sec:acid}). In order to represent the two interacting fluids discretely, the indicator function $\zeta$ translates into a colour function $\psi$, defined for cell $P$ as
\begin{equation}
\psi_P = \frac{1}{V_P} \int_{V_P} \zeta \, \mathrm{d}V.
\label{eq:colour}
\end{equation}
The interface is, thus, located in every cell with $0 < \psi < 1$.

\subsection{Interface advection}
\label{sec:interfaceAdvection}
To advect the fluid interface between two fluids, Eq.~(\ref{eq:vofAdvection}) is applied to the colour function, Eq.~(\ref{eq:colour}), and reformulated as
\begin{equation}
\frac{\partial \psi}{\partial t} + \mathbf{\nabla} \cdot (\mathbf{u} \psi) - \psi \, \mathbf{\nabla} \cdot \mathbf{u} = 0. \label{eq:vofAdvComp}
\end{equation} 
The advection of the fluid interface is then described by Eq.~(\ref{eq:vofAdvComp}) using an appropriate discretisation method. Two different VOF methods, an algebraic VOF method \cite{Denner2018b} and a piecewise-linear interface calculation (PLIC) method with Lagrangian advection of the interface \cite{vanWachem2002}, are considered for the advection of the fluid interface. 

In the algebraic VOF method \cite{Denner2018b}, the advection equation of the colour function, Eq.~(\ref{eq:vofAdvComp}), is discretised using the Crank-Nicolson scheme for the discretisation of the transient term and the CICSAM scheme \cite{Ubbink1999} for the spatial interpolation of the colour function $\psi$. 
In the applied VOF-PLIC method, the interface is reconstructed based on the local colour function $\psi$ and the normal vector of the interface \cite{Scardovelli2000}. The interface advection equation (\ref{eq:vofAdvComp}) is then rewritten in integral form and the reconstructed interface is advected using the Lagrangian split advection scheme of van Wachem and Schouten \cite{vanWachem2002}.
In both considered interface advection methods, the advection is based on the same advecting velocity $\vartheta_f$ as for all advection terms of the governing equations, thus ensuring an accurate prediction of volume changes \cite{Denner2018b}.

Nevertheless, the finite-volume discretisation and pressure-based algorithm presented in Section \ref{sec:numericalFramework} are not limited to the employed VOF methods and other methods to represent the bulk phases and advect the interface, including level-set and front-tracking methods, may equally be applied.

\subsection{Fluid properties}
\label{sec:fluidProp}
The definitions of the fluid properties largely follow the principles outlined by Denner et al.~\cite{Denner2018b}. The density of the fluid is defined based on the colour function $\psi$ as
\begin{equation}
\rho = (1 - \psi) \, \rho_\mathrm{a} +  \psi \,
\rho_\mathrm{b} , \label{eq:densityFluid}
\end{equation}
where the partial densities $\rho_\mathrm{a}$ and $\rho_\mathrm{b}$ of the bulk phases are given by Eq.~(\ref{eq:rhoFull}). This linear interpolation of the density is required in order to satisfy the discrete conservation of mass, momentum and energy. In the context of compressible flows, it is equivalent to an isobaric closure assumption \cite{Allaire2002,Shyue2006}, while for incompressible fluids this formulation reduces to the typically used definition of density \cite{Brackbill1992}. The heat capacity ratio also follows from the isobaric closure assumption as
\begin{equation}
\frac{1}{\gamma - 1} = \mathcal{C}_\mathrm{a} \,
\frac{1-\psi}{\gamma_{0,\mathrm{a}} - 1} + \mathcal{C}_\mathrm{b} \,
\frac{\psi}{\gamma_{0,\mathrm{b}} - 1} ,
\label{eq:gammaMinusOne}
\end{equation}
where $\mathcal{C}_\mathrm{a}$ and $\mathcal{C}_\mathrm{b}$ are the binary compressibility operators defined in Eq.~(\ref{eq:operatorC}) associated with fluid $\mathrm{a}$ and fluid $\mathrm{b}$, respectively.
The specific isobaric heat capacity is defined by a mass-weighted interpolation \cite{Denner2018b}, which is essential for the conservation of the total energy, given as
\begin{equation}
\rho \, c_p = (1 - \psi) \, \rho_\mathrm{a} \, c_{p,\mathrm{a}} 
+ \psi \, \rho_\mathrm{b} \, c_{p,\mathrm{b}} ,
\label{eq:cpFluid}
\end{equation}
where the partial densities $\rho_\mathrm{a}$ and $\rho_\mathrm{b}$ are given by Eq.~(\ref{eq:rhoFull}), density $\rho$ is given by Eq.~(\ref{eq:densityFluid}), and the partial specific isobaric heat capacities $c_{p,\mathrm{a}}$ and $c_{p,\mathrm{b}}$ are given by Eq.~(\ref{eq:cpFull}). The viscosity $\mu$ and thermal conductivity $k$ are defined as
\begin{align}
\mu & = (1 - \psi) \, \mu_\mathrm{a} +  \psi \, \mu_\mathrm{b} , \\
k & = (1 - \psi) \, k_\mathrm{a} +  \psi \, k_\mathrm{b} .
\end{align}

The {\em Continuum Surface Force} (CSF) model of Brackbill et al.~\cite{Brackbill1992} is applied to model the force due to surface tension,
represented by the source term 
\begin{equation}
\mathbf{S} = \sigma \, \kappa \, \mathbf{\nabla} \psi , \label{eq:csfFull}
\end{equation}
where $\kappa$ is the interface curvature, which is computed using a second-order height-function method \cite{Evrard2020}, and $\sigma$ is the surface tension coefficient. In the interest of conciseness, but without loss of generality, the surface tension coefficient is assumed to be constant.

\subsection{Coupling of the bulk phases}
\label{sec:acid}
The ACID method \cite{Denner2018b} assumes that, for the purpose of discretising the governing conservation laws for a given cell, all cells in its finite-volume stencil are assigned the same colour function value, {\em i.e.}~the colour function is assumed to be constant in the entire finite-volume stencil. The relevant thermodynamic properties that are discontinuous at the interface, {\em i.e.}~density and enthalpy, are then evaluated based on this locally constant colour function field, as described by Denner et al.~\cite{Denner2018b} in detail. This recovers the contact discontinuity associated with the interface \cite{Anderson2003,Toro2009} and enables the application of the conservative discretisation described in Section \ref{sec:numericalFramework}, identical to the one applied for single-phase flows. Denner et al.~\cite{Denner2018b} reported robust and accurate results for acoustic and shock waves in interfacial flows, supporting the notion that the interface discretisation indeed conserves the acoustic features of the flow and retains the conservative discretisation of the governing equations.

\subsubsection{Density treatment}
Under the assumption that the colour function $\psi$ is constant throughout the finite-volume stencil of cell $P$, as proposed by Denner et al.~\cite{Denner2018b}, the density interpolated to face $f$ of cell $P$ is given as
\begin{equation}
{\rho}_f =  \rho_U^\star  + \frac{\xi_{f}}{2} \left( \rho_D^\star - \rho_U^\star \right) , \label{eq:densityFaceDiscNew}
\end{equation} 
where $\xi_f$ is the flux limiter determined by the applied differencing scheme, {\em e.g.}~a TVD differencing scheme \cite{Denner2015a}.
The density $\rho_U^\star$ at the upwind cell $U$ and $\rho_D^\star$ at the downwind cell $D$ are given based on the colour function value of cell $P$ by Eq.~(\ref{eq:densityFluid}), so that
\begin{equation}
\rho_U^\star = \rho_{\mathrm{a},U} +  \psi_P \, \left(\rho_{\mathrm{b},U} -
\rho_{\mathrm{a},U} \right)  \label{eq:rhoUstar}
\end{equation} 
and
\begin{equation}
\rho_D^\star = \rho_{\mathrm{a},D} + 
\psi_P \, \left(\rho_{\mathrm{b},D} -
\rho_{\mathrm{a},D} \right) .  \label{eq:rhoDstar}
\end{equation}
The densities at previous time-levels are evaluated in a similar fashion based on the colour function value of cell $P$, with
\begin{equation}
\rho_P^{(t-\Delta t)} = \rho_{\mathrm{a},P}^{(t-\Delta t)} + 
\psi_P^{(t)} \, \left(\rho_{\mathrm{b},P}^{(t-\Delta t)} -
\rho_{\mathrm{a},P}^{(t-\Delta t)}\right) \label{eq:acidDensityT1}
\end{equation}
and analogously for $\rho_P^{(t-2 \Delta t)}$.

\subsubsection{Enthalpy treatment}
The specific total enthalpy at face $f$ is given, again assuming the colour function $\psi$ is constant throughout the finite-volume stencil of
cell $P$, as \cite{Denner2018b}
\begin{equation}
{\rho}_f {h}_f = \rho_U^\star h_U^\star +
\frac{\xi_{f}}{2} \left(\rho_D^\star h_D^\star - \rho_U^\star h_U^\star
\right) , \label{eq:enthalpyACIDCorr}
\end{equation}
with ${\rho}_f$ given by Eq.~(\ref{eq:densityFaceDiscNew}). The specific total enthalpy of the upwind and downwind cells are 
\begin{align}
h_U^\star & = c_{p,U}^\star \, T_U + \frac{\mathbf{u}^{2}_U}{2}  , \\
h_D^\star & = c_{p,D}^\star \, T_D + \frac{\mathbf{u}^{2}_D}{2}  ,
\end{align}
respectively, $\rho_U^\star$ is given by Eq.~(\ref{eq:rhoUstar}) and $\rho_D^\star$ is given by Eq.~(\ref{eq:rhoDstar}). The specific isobaric heat capacities $c_{p,U}^\star$ and $c_{p,D}^\star$ are defined by Eq.~(\ref{eq:cpFluid}) with $\psi_P$ as
\begin{equation}
\rho_U^\star \, c_{p,U}^\star = \rho_{\mathrm{a},U} \, c_{p,\mathrm{a},U} + \psi_P \, (\rho_{\mathrm{b},U} \, c_{p,\mathrm{b},U} - \rho_{\mathrm{a},U} \,
c_{p,\mathrm{a},U})
\end{equation}
and
\begin{equation}
\rho_D^\star \, c_{p,D}^\star = \rho_{\mathrm{a},D} \, c_{p,\mathrm{a},D} + \psi_P \, (\rho_{\mathrm{b},D} \, c_{p,\mathrm{b},D}-\rho_{\mathrm{a},D} \,
c_{p,\mathrm{a},D}) .
\end{equation}
Since the specific enthalpy is partially sought implicitly, formulated implicit with respect to the primary solution variable temperature, see Eq.~(\ref{eq:enthalpyImplemented}), a deferred correction approach is applied to enforce Eq.~(\ref{eq:enthalpyACIDCorr}) \cite{Denner2018b}.

The specific total enthalpy at the previous time-levels is given as
\begin{equation}
h_P^{(t-\Delta t)} = c_{p,P}^{\star,(t-\Delta t)} \, T_P^{(t-\Delta t)} + \dfrac{\mathbf{u}^{(t-\Delta t),2}_P}{2} ,
\end{equation}
with
\begin{equation}
\begin{split}
\rho_{P}^{(t-\Delta t)} \, c_{p,P}^{\star,(t-\Delta t)}  = \rho_{\mathrm{a},P}^{(t-\Delta t)} \, c_{p,\mathrm{a},P}^{(t-\Delta t)}  + \ \psi_P^{(t)} \left(\rho_{\mathrm{b},P}^{(t-\Delta t)} \, c_{p,\mathrm{b},P}^{(t-\Delta t)} - \rho_{\mathrm{a},P}^{(t-\Delta t)} \, c_{p,\mathrm{a},P}^{(t-\Delta t)}\right) ,
\end{split}
\end{equation}
and analogously for $h_P^{(t- 2 \Delta t)}$ and $c_{p,P}^{\star,(t-2 \Delta t)}$.

\subsubsection{Further observations}
The corrections applied by ACID to the discretised governing equations vanish in the bulk phases and are non-zero only at fluid interfaces \cite{Denner2018b}. The fluid properties are piecewise constant at the interface and the corresponding error is proportional to $\Delta x$, as commonly found in VOF, level-set and front-tracking methods \cite{Popinet2018}. With respect to the different fluid combinations that can occur at the interface, the following can be observed:
\begin{itemize}
\item At compressible-compressible interfaces, problems associated with a discontinuous change of fluid properties are circumvented with ACID, while retaining the information carried by compression and expansion waves, as comprehensively demonstrated by Denner et al.~\cite{Denner2018b}. Hence, acoustic waves, shock waves and rarefaction fans can interact with and at the interface. The proposed algorithm then becomes an enhanced version of the algorithm for compressible interfacial flows of Denner et al.~\cite{Denner2018b}, including viscous stresses and surface tension.

\item At incompressible-incompressible interfaces, ACID retains the incompressible formulation of the discretised governing equations. In fact, the non-conservative formulation of the governing equations at the interface originally proposed by Brackbill et al.~\cite{Brackbill1992} is obtained. The proposed algorithm then reduces to a non-isothermal version of the fully-coupled algorithm for incompressible interfacial flows of Denner and van Wachem \cite{Denner2014a}, as for instance used in \cite{Denner2015}.

\item At compressible-incompressible interfaces, the compressibility of the compressible fluid stays, dependent on the local colour function value, partly intact and ACID provides a transition from the compressible fluid to the incompressible fluid. Compression and expansion waves are able to interact with such a compressible-incompressible interface, but compressible effects are not transmitted into the incompressible fluid.
\end{itemize}

\section{Validation}
\label{sec:results}
As already discussed, for incompressible-incompressible interfacial flows, the proposed algorithm is equivalent to the incompressible algorithm of Denner and van Wachem \cite{Denner2014a}, which has been extensively tested and validated against analytical solutions \cite{Denner2015, Denner2017a}, experiments \cite{Denner2018, Denner2014e} and other numerical methods \cite{Denner2017a}. For compressible-compressible interfacial flows, the proposed algorithm becomes the compressible algorithm of Denner et al.~\cite{Denner2018b}, which has been validated thoroughly with a broad range of test-cases. Thus, the validation presented below focuses on the application of the proposed algorithm to predict compressible-incompressible interfacial flows. Five representative test-cases are considered to demonstrate the application of the proposed algorithm to compressible-incompressible interfacial flows: a bubble with surface tension in equilibrium (Section \ref{sec:bubbleEquilibrium}), the viscous damping of capillary waves (Section \ref{sec:capillaryWaves}), the reflection of an acoustic wave at a fluid interface (Section \ref{sec:acousticWaves}), the pressure-driven collapse of a bubble (Section \ref{sec:bubbleCollapse}) and the shock interaction with a water drop (Section \ref{sec:shockDrop}). 

\subsection{Bubble in equilibrium}
\label{sec:bubbleEquilibrium}

A circular bubble of a compressible gas surrounded by an incompressible fluid is simulated to study the parasitic currents associated with the numerical treatment of the surface tension contribution. Similar to the test-cases used in previous studies \cite{Popinet2009,Fuster2018}, the considered two-dimensional bubble has a diameter of $d_\mathrm{B} = 0.8 \, \mathrm{m}$ and a surface tension coefficient of $\sigma = 1.0 \, \mathrm{N} \, \mathrm{m}^{-1}$. The ambient pressure is $p_0 = 1.0 \, \mathrm{Pa}$, the incompressible fluid has a density of $\rho_{0,\mathrm{f}} = 1.0 \, \mathrm{kg} \, \mathrm{m}^{-3}$ and the compressible gas has a density of $\rho_{\mathrm{g}} = 1.0 \, \mathrm{kg} \, \mathrm{m}^{-3}$ at $p_0$. The compressible gas has a heat capacity ratio of $\gamma_0 = 1.4$ and both fluids have a specific isobaric heat capacity of $c_{p,0} = 1008 \, \mathrm{J} \, \mathrm{kg}^{-1} \, \mathrm{K}^{-1}$. The viscosity $\mu$, which is the same for both fluids, follows from the considered Laplace number, $\mathrm{La} = \rho_{0,\mathrm{f}}  d_\mathrm{B}  \sigma/\mu^2 = 120$. The bubble is placed at the centre of a domain with edge length $2 \, \mathrm{m} \times 2 \, \mathrm{m}$, represented by an equidistant Cartesian mesh with $64 \times 64$ mesh cells, and the applied time-step is $\Delta t = 10^{-3} \, \mathrm{s}$, which satisfies the capillary time-step constraint \cite{Denner2015}. The fluid interface is advected using the VOF-PLIC method.

\begin{figure}[t]
\begin{center}
\includegraphics[width=0.45\textwidth]{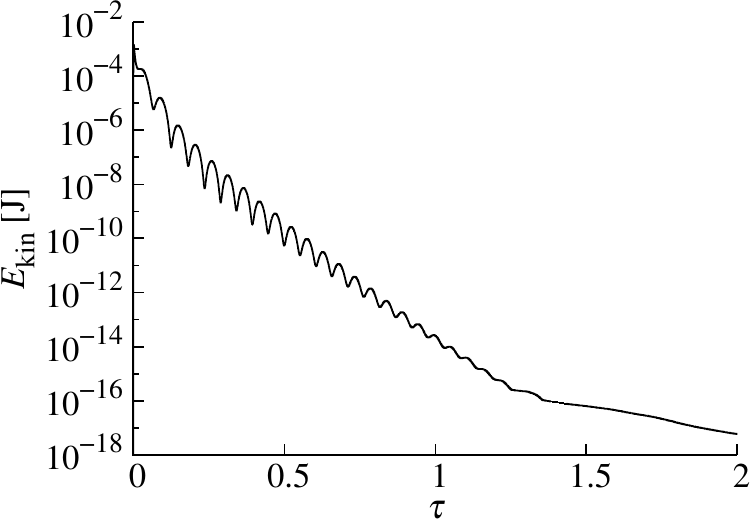}
\caption{Kinetic energy $E_\mathrm{kin}$ as a result of parasitic currents as a function of dimensionless time $\tau = t/t_\mu$, with $t_\mu = \rho_{0,\mathrm{f}} d_\mathrm{B}^2/\mu$ the viscous timescale, of the bubble in equilibrium. The gas inside the bubble is compressible, whereas the fluid surrounding the bubble is incompressible.}
\label{fig:bubbleEquilibrium}
\end{center}
\end{figure}

Figure \ref{fig:bubbleEquilibrium} shows the total kinetic energy $E_\mathrm{kin} = \sum_{P=1}^N \rho_P \mathbf{u}_P^2/2$, with $N$ the total number of mesh cells, in the computational domain as a function of the dimensionless time $\tau = t/t_\mu$, with $t_\mu = \rho_{0,\mathrm{f}} d_\mathrm{B}^2/\mu$ the viscous timescale. Since the bubble is in equilibrium, the observed kinetic energy is the result of parasitic currents only. The imbalance caused by errors in the numerical evaluation of the interface curvature leads to an initial production of parasitic currents and the associated kinetic energy. As the interface topology relaxes towards a numerical equilibrium, the production of parasitic currents diminishes and the existing parasitic currents dissipate as a result of viscous stresses. Consequently, the kinetic energy in the domain decays rapidly and the exact balance is recovered on the discrete level.

\subsection{Capillary waves}
\label{sec:capillaryWaves}

To demonstrate the accurate prediction of viscous flows and surface tension by the proposed algorithm, the oscillation of a standing capillary wave between two viscous fluids is simulated. This test case is particularly challenging because the wave amplitude is small and the temporal evolution of the wave amplitude, governed by the dispersion (due to surface tension) and attenuation (due to viscous stresses) of the capillary wave, is very sensitive to the implementation of the viscous stress terms, the surface tension model, numerical diffusion and spurious oscillations \cite{Denner2017a, Popinet2009}. The analytical solution for the initial value problem of a freely-oscillating capillary wave with infinitesimal amplitude, derived by Prosperetti \cite{Prosperetti1981}, which is valid for two-phase systems in which both fluids have the same kinematic viscosity $\nu = \mu/\rho$ or in which one fluid is neglected, serves as a reference. The considered capillary wave has a wavelength of $\lambda = 0.1 \, \mathrm{m}$, both fluids have a kinematic viscosity of $\nu = 1.6394 \times 10^{-4} \, \mathrm{m}^2 \, \mathrm{s}^{-1}$ and uniform initial temperature $T_0 = 300 \, \mathrm{K}$, and the surface tension coefficient is $\sigma = 0.25 \, \pi^{-3} \, \mathrm{N} \, \mathrm{m}^{-1}$. The density of the two fluids are $\rho_\mathrm{a} = 1 \, \mathrm{kg} \, \mathrm{m}^{-3}$ and $\rho_\mathrm{b} = 16 \, \mathrm{kg} \, \mathrm{m}^{-3}$. The computational domain has the dimensions $\lambda \times 3\lambda$ and is represented by an equidistant Cartesian mesh with $\Delta x = \lambda / 32$. A compressible-compressible flow, with $\gamma_0 = 1.4$, as well as a compressible-incompressible flow (fluid 'b' is considered to be incompressible) are considered.  The fluid interface is advected using the algebraic VOF method.

\begin{figure}[t]
\begin{center}
\subfloat[Compressible-compressible flow]
{\includegraphics[width=0.47\textwidth]{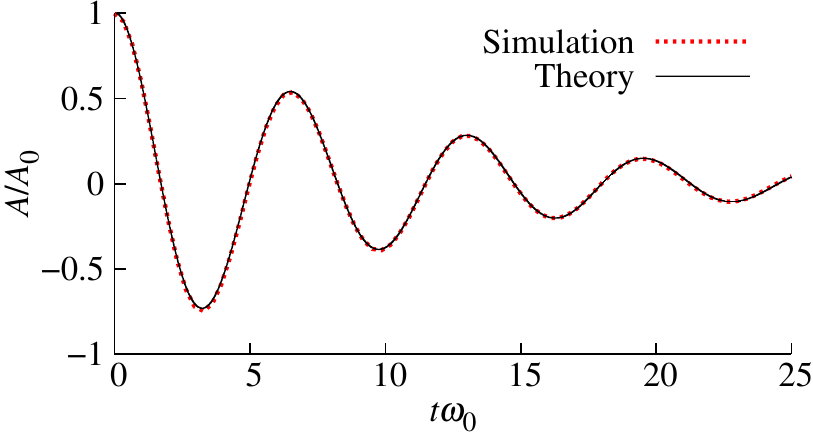}}
\quad
\subfloat[Compressible-incompressible flow]
{\includegraphics[width=0.47\textwidth]{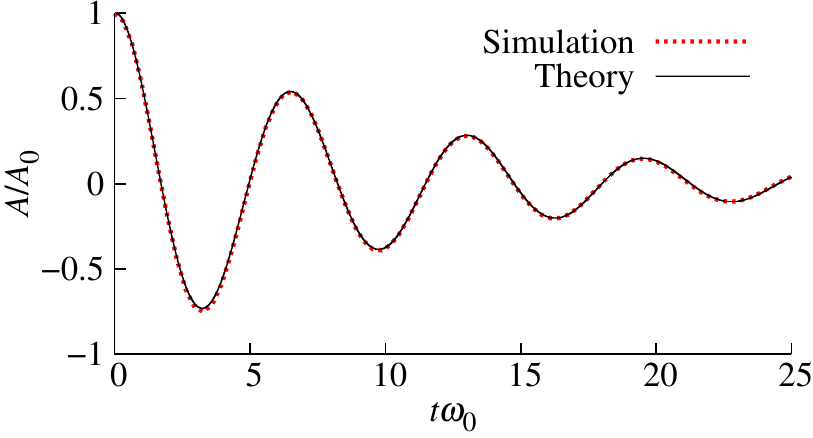}}
\caption{Temporal evolution of the dimensionless amplitude $A/A_0$ of a capillary wave with wavelength $\lambda = 0.1 \, \mathrm{m}$ and initial amplitude $A_0 = 0.01 \lambda$ as a function of dimensionless time $t \omega_0$, obtained for a compressible-compressible flow and a compressible-incompressible flow. The results are compared against the analytical solution of Prosperetti \cite{Prosperetti1981}.}
\label{fig:capillaryWaveComp0p01}
\end{center}
\end{figure}

Figure \ref{fig:capillaryWaveComp0p01} shows the dimensionless wave amplitude $A/A_0$ as a function of the dimensionless time $t \omega_0$, where $\omega_0 = \sqrt{\sigma k^3/(\rho_\mathrm{a} + \rho_\mathrm{b})}$ is the undamped angular frequency of the capillary wave and $k=2\pi/\lambda$ is the wavenumber, with initial wave amplitude $A_0 = 0.01\lambda$, obtained for both flow types. In both cases the computed result is in excellent agreement with the analytical solution of Prosperetti \cite{Prosperetti1981}, with respect to the amplitude as well as the frequency.

\subsection{Acoustic waves}
\label{sec:acousticWaves}
The propagation of a single acoustic wave in an air-water flow, where air is treated as a compressible fluid and water is taken to be either a compressible fluid described by the stiffened-gas model or an incompressible fluid with constant density, is simulated in a one-dimensional domain with mesh spacing $\Delta x = 2 \times 10^{-3} \, \mathrm{m}$, with   initial velocity $u_0 = 1 \, \mathrm{m} \, \mathrm{s}^{-1}$, initial pressure $p_0 = 10^5 \, \mathrm{Pa}$ and initial temperature $T_0=300 \, \mathrm{K}$. Air has the fluid properties $\gamma_\mathrm{0,Air} = 1.4$ and $\Pi_\mathrm{0,Air} = 0  \, \mathrm{Pa}$, and the density at initial conditions is $\rho_\mathrm{Air} = 1.16 \, \mathrm{kg} \, \mathrm{m}^{-3}$. Compressible water has the fluid properties $\gamma_\mathrm{0,Water} = 4.1$, $\Pi_\mathrm{0,Water} = 4.4 \times 10^8 \, \mathrm{Pa}$ with a density at initial conditions of $\rho_\mathrm{Water} = 1000 \, \mathrm{kg} \, \mathrm{m}^{-3}$.  Incompressible water has a {\em constant} density of $\rho_\mathrm{0,Water} = 1000 \, \mathrm{kg} \, \mathrm{m}^{-3}$. The acoustic wave is initiated by the inlet-velocity
\begin{equation}
u_\mathrm{in} =
\begin{cases}
u_0 + \Delta u_0 \, \sin \left(2 \pi f t + \dfrac{3}{2}
\pi\right) & \mbox{if $t < f^{-1}$}
\\
u_0 - \Delta u_0 & \mbox{if $t \geq f^{-1}$} ,
\end{cases}
\end{equation}
with $\Delta u_0$ the amplitude and $f=2500 \, \mathrm{s}^{-1}$ the frequency of the velocity perturbation. Two perturbation amplitudes are considered, $\Delta u_0 \in \{0.01 \, u_0, 100 \, u_0\}$, to study the linear and nonlinear acoustic regimes. 

In the linear acoustic regime ($\Delta u \ll a_0$ and $\Delta \rho \ll \rho_0$), which applies to the considered velocity perturbation amplitude of $\Delta u_0 = 0.01 \, u_0$, the perturbation leads to a sound wave. Based on linear acoustic theory \cite{Anderson2003}, the acoustic wave reflected at the air-water interface should have a pressure amplitude of
\begin{equation}
\Delta p_{\mathrm{Air},0}^\mathrm{reflected} = \dfrac{\Delta
p_{\mathrm{Air},0}^\mathrm{incident}}{\dfrac{2
Z_\mathrm{Water}}{Z_\mathrm{Water} - Z_\mathrm{Air}} - 1} ,
\end{equation}
where $Z = \rho a$ is the characteristic acoustic impedance and $\Delta p_{\mathrm{Air},0}^\mathrm{incident} =  Z_\mathrm{Air} \, \Delta u_0$ is the pressure amplitude of the incident wave. While the theoretical pressure amplitude of the reflected wave is $\Delta p_{\mathrm{Air},0}^\mathrm{reflected} = \Delta p_{\mathrm{Air},0}^\mathrm{incident}$ if water is considered to be an incompressible fluid ($\rho_\mathrm{Water} = \mathrm{const.}$, $a_\mathrm{Water} = \infty$), the reflected pressure amplitude is $\Delta p_{\mathrm{Air},0}^\mathrm{reflected} \simeq 0.999 \, \Delta p_{\mathrm{Air},0}^\mathrm{incident}$ if water is considered to be a compressible fluid ($\rho_\mathrm{Water} = f(p,T)$, $a_\mathrm{Water} = 1343 \, \mathrm{m} \, \mathrm{s}^{-1}$). Hence, the differences in the results between the incompressible and compressible treatments of water should be negligible, at least in the linear acoustic regime, apart from the transmitted wave that should only be present for the compressible treatment of water. 

\begin{figure}[t]
\begin{center}
\subfloat[$\Delta u_0 = 0.01 \, u_0$]{
\includegraphics[width=0.49\textwidth]{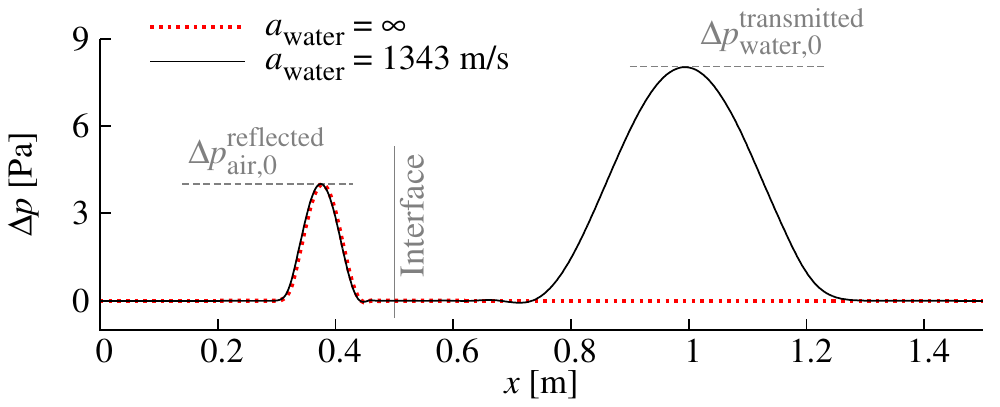}}
\hfill
\subfloat[$\Delta u_0 = 100 \, u_0$]{
\includegraphics[width=0.49\textwidth]{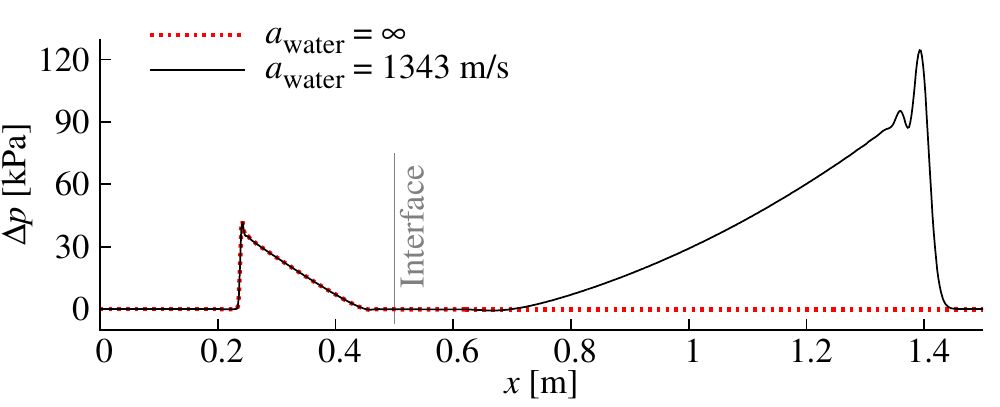}}
\caption{Pressure profile of acoustic waves with different amplitudes in a one-dimensional domain after the interaction with an air-water interface at $t=2 \, \mathrm{ms}$, in which air (left of the interface) is treated as a compressible fluid and water (right of the interface) is treated as an incompressible fluid ($a_\mathrm{Water} = \infty$) or as a compressible fluid ($a_\mathrm{Water} = 1343 \, \mathrm{m} \, \mathrm{s}^{-1}$). The amplitude of the applied initial velocity perturbation is (a) $\Delta u_0 = 0.01 \, u_0$ and (b) $\Delta u_0 = 100 \, u_0$. The pressure amplitudes of the reflected and transmitted waves according to linear acoustic theory are shown in (a) as a reference.}
\label{fig:incompCompAcoustic}
\end{center}
\end{figure}

Figure \ref{fig:incompCompAcoustic} shows the pressure profile of the acoustic waves, after they have interacted with the interface, for both considered perturbation amplitudes using both the compressible treatment and the incompressible treatment of the liquid phase. For the small perturbation amplitude, $\Delta u_0 = 0.01 \, u_0$, the amplitude of the reflected pressure wave is in excellent agreement with linear acoustic theory with both treatments of the liquid phase. When the liquid phase is treated as a compressible fluid, a pressure wave is transmitted through the interface, while no wave is transmitted when the liquid phase is treated as an incompressible fluid, as expected.
For the large perturbation amplitude, $\Delta u_0 = 100 \, u_0$, the pressure profile of the acoustic wave departs significantly from its originally sinusoidal shape as a result of nonlinear wave steepening. Nevertheless, the reflected waves predicted by the compressible treatment and the incompressible treatment of the liquid phase are in excellent agreement, suggesting that the compressible-incompressible treatment of gas-liquid flows can be applied even when large amplitude acoustic waves are present in the gas phase.

\subsection{Bubble collapse}
\label{sec:bubbleCollapse}
The pressure-driven growth and collapse of gas bubbles, in particular cavitation, is a widely observed phenomenon, and the ability to focus large amounts of energy by a collapsing bubble is being utilised in an increasing number of applications. The simplest analytical model for the pressure-driven collapse of a bubble in a liquid is the Rayleigh-Plesset equation \cite{Plesset1949} without considering viscous dissipation and surface tension, given as
\begin{equation}
R \ddot{R} + \frac{3}{2} \dot{R}^2 = \frac{p_\mathrm{g} - p_\infty}{\rho_{0,\ell}},
\label{eq:RP}
\end{equation}
where $R$ is the bubble radius, $\rho_{0,\ell}$ is the (constant) density of the liquid,
$p_\mathrm{g}$ is the (uniform) gas pressure inside the bubble and $p_\infty$ is the ambient liquid pressure at infinite distance from the bubble. Interestingly, Eq.~(\ref{eq:RP}) is based on the assumption of an incompressible liquid, with $\rho_{0,\ell} = \mathrm{const.}$, and a compressible gas bubble.

The collapse of a spherical gas bubble due to an overpressure in the liquid, the so-called {\em Rayleigh collapse} \cite{Lauterborn2018}, is considered to validate the prediction of pressure-driven flows of a compressible gas in contact with an incompressible liquid by the proposed algorithm.
Following the setup of Denner et al.~\cite{Denner2020a}, a bubble with initial radius $R_0$ and an initial gas pressure of $p_\mathrm{g} = 4000 \, \mathrm{kPa}$ is situated in a liquid with an initial pressure of $p_\ell(r) = p_\infty + (p_\mathrm{g}-p_\infty) \, R_0/r $, with $p_\infty = 10^5 \, \mathrm{Pa}$ and $r$ the radial coordinate. The gas has a heat capacity ratio of $\gamma_0 = 1.4$  and the liquid has a constant density of $\rho_{0,\ell} = 1000 \, \mathrm{kg} \, \mathrm{m}^{-3}$. Viscosity, thermal conduction and surface tension are neglected. The fluid interface is advected using the algebraic VOF method.
Because the liquid is incompressible and viscous dissipation and thermal conduction are neglected, the oscillations of the bubble radius should continue indefinitely with unchanged frequency and amplitude. 

\begin{figure}[t]
  \begin{center}
  \subfloat[$\Delta x = R_0/400$]
  {\includegraphics[width=0.4\textwidth]{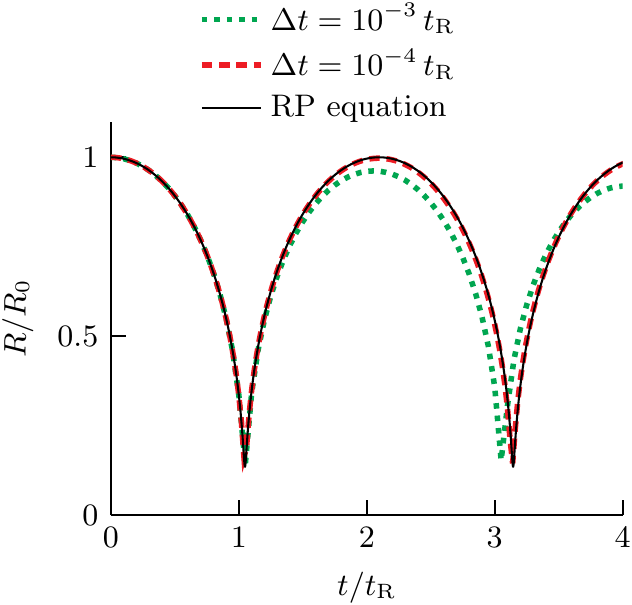}}
  \quad
  \subfloat[$\Delta t = 10^{-4} t_\mathrm{R}$]
  {\includegraphics[width=0.4\textwidth]{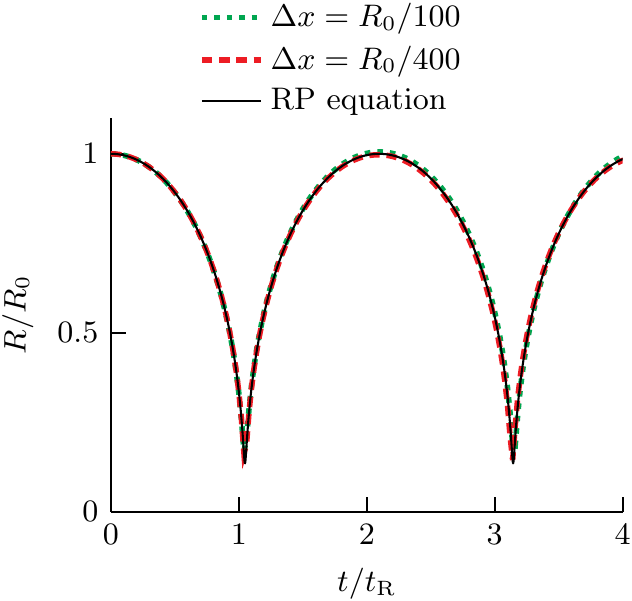}}
  \caption{Temporal evolution of the dimensionless radius $R/R_0$ of a collapsing gas bubble in an incompressible liquid, compared against the solution of the Rayleigh-Plesset equation (\ref{eq:RP}). (a) Results for different time-steps $\Delta t$ obtained with a spatial resolution of $\Delta x = R_0/400$; (b) Results with different mesh spacings $\Delta x$ obtained with a time-step of $\Delta t = 10^{-4} \, t_\mathrm{R}$. Time $t$ is normalised by the Rayleigh collapse time, $t_\mathrm{R} = 0.915 \, R_0 \sqrt{\rho_{0,\ell}/p_\infty}$.}
  \label{fig:bubbleCollapse}
  \end{center}
  \end{figure}

Figure \ref{fig:bubbleCollapse} shows the dimensionless radius $R/R_0$ as a function of the dimensionless time $t/t_\text{R}$, where $t_\text{R} = 0.915 \, R_0 \sqrt{\rho_{0,\ell}/p_\infty}$ is the Rayleigh collapse time, obtained with different spatial and temporal resolutions using the proposed algorithm. The solution of the Rayleigh-Plesset equation (\ref{eq:RP}) is also shown as a reference and regarded here as the {\em exact} solution of the dynamic bubble behaviour under consideration.
The results predicted by the proposed algorithm and by the Rayleigh-Plesset equation are in excellent agreement, as shown in Fig.~\ref{fig:bubbleCollapse}, for a sufficiently large spatial and temporal resolution.

\subsection{Shock-drop interaction}
\label{sec:shockDrop}

\begin{figure}[t]
  \begin{center}
  \subfloat[Velocity]
  {\includegraphics[width=0.47\textwidth]{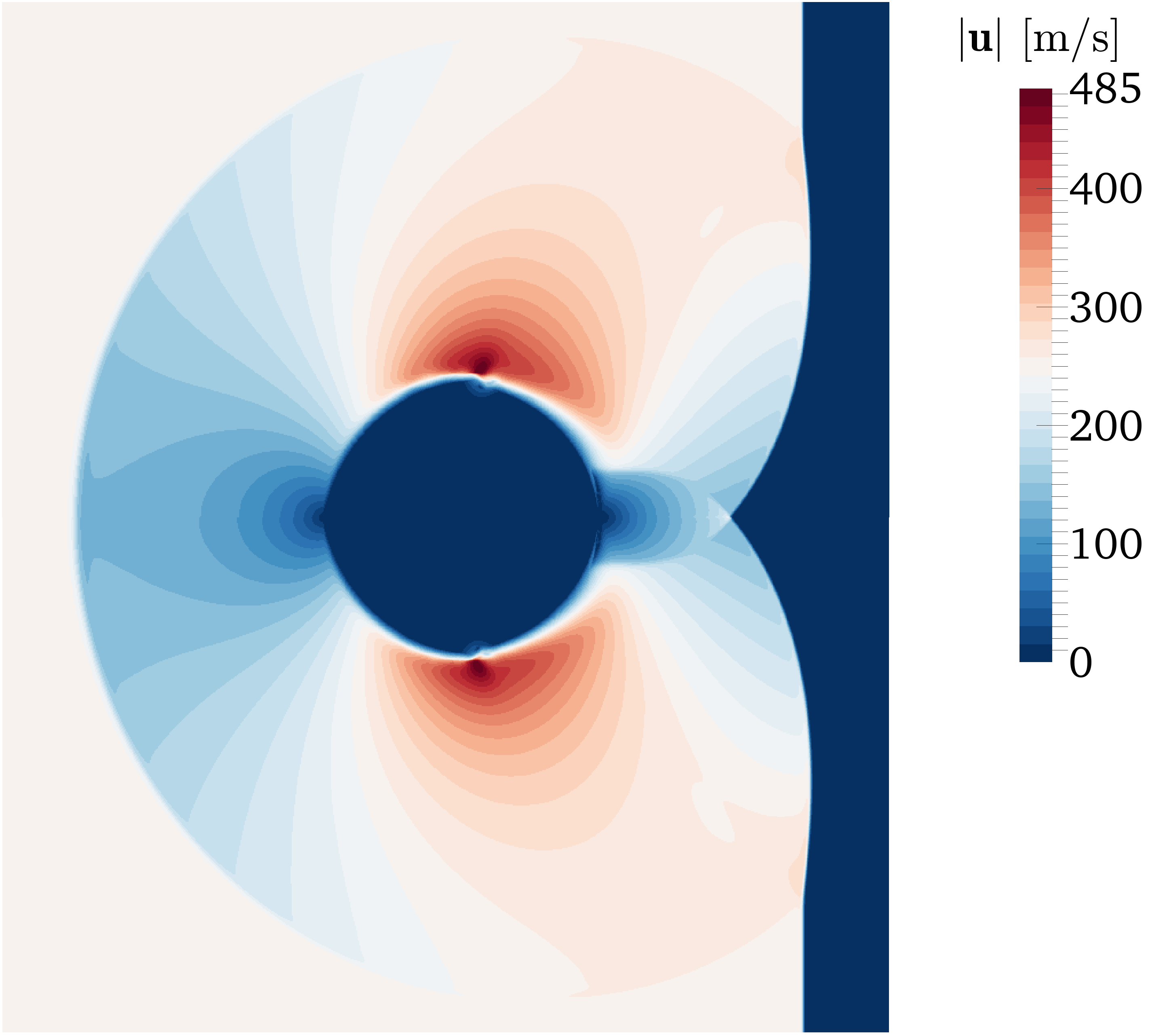}}
  \hfill
  \subfloat[Density gradient]
  {\includegraphics[width=0.47\textwidth]{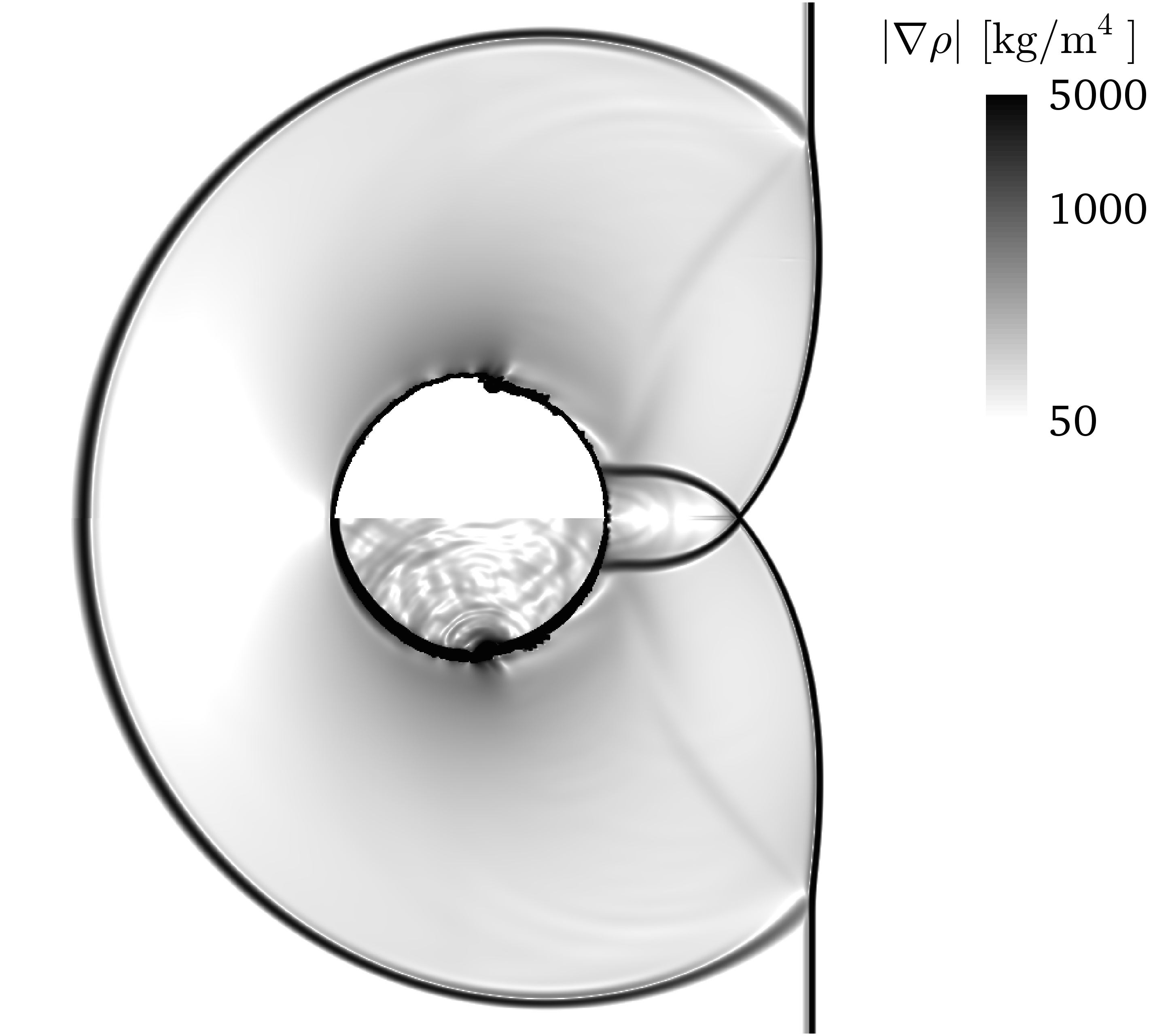}}
  \caption{Contours of the velocity magnitude $|\mathbf{u}|$ and the density gradient $|\mathbf{\nabla} \rho|$ after a shock with $M_\mathrm{s} = 1.47$ interacted with a two-dimensional water drop, simulating the water drop as an incompressible fluid (upper half) and as a compressible fluid (lower half). The density waves inside the compressible drop are clearly visible, while no density waves are transmitted to the incompressible drop.}
  \label{fig:shockDropAPS}
  \end{center}
  \end{figure}

Following the work of Meng and Colonius \cite{Meng2015}, the interaction of a shock wave, initially travelling in air with Mach number $M_\mathrm{s} = 1.47$, with a water drop is simulated. The two-dimensional domain is represented by an equidistant Cartesian mesh with $100$ cells per drop diameter. The shock wave separates the post-shock region (I) and the pre-shock region (II), which are initialised with $u_\mathrm{I} = 246.24 \, \mathrm{m} \, \mathrm{s}^{-1}$, $p_\mathrm{I} = 2.35 \times 10^5 \, \mathrm{Pa}$, $T_\mathrm{I} = 450.56 \, \mathrm{K}$ and $u_\mathrm{II} = 0  \, \mathrm{m} \, \mathrm{s}^{-1}$, $p_\mathrm{II} = 1.00 \times 10^5 \, \mathrm{Pa}$, $T_\mathrm{II} = 347.22 \, \mathrm{K}$, respectively.  The fluid properties of air are $\gamma_{0,\mathrm{Air}} = 1.4$ and $\Pi_{0,\mathrm{Air}} = 0  \, \mathrm{Pa}$, with $\rho_\mathrm{II,Air} = 1 \, \mathrm{kg} \, \mathrm{m}^{-3}$ and $a_\mathrm{II,Air} = 374.17 \, \mathrm{m} \, \mathrm{s}^{-1}$ in the pre-shock region. The liquid drop is treated as an incompressible fluid, with constant density $\rho_\mathrm{0,Water} = 1000 \, \mathrm{kg} \, \mathrm{m}^{-3}$, or as a compressible fluid, with the fluid properties  $\gamma_{0,\mathrm{Water}} = 4.4$ and $\Pi_{0,\mathrm{Water}} = 6 \times 10^8 \, \mathrm{Pa}$, and $\rho_\mathrm{II,Water} = 1000 \, \mathrm{kg} \, \mathrm{m}^{-3}$ and $a_\mathrm{II,Water} = 1624.94 \, \mathrm{m} \, \mathrm{s}^{-1}$ in the pre-shock region. The fluid interface is advected using the algebraic VOF method.

Figure \ref{fig:shockDropAPS} shows the contours of the velocity magnitude $|\mathbf{u}|$ and the density gradient $|\mathbf{\nabla}\rho|$ obtained $t=16 \, \mu \mathrm{s}$ after the first shock-drop interaction, treating water as an incompressible or a compressible fluid. While the velocity profile and the shock structures are virtually identical in the gas phase, the compressible treatment of the water drop allows density waves to propagate in it, while no density waves are propagating in the incompressible drop. With respect to the execution time of the simulations, the incompressible treatment of the drop achieves a significant speed-up of the simulation, because the stiff pressure-density coupling imposed by the stiffened-gas model for the compressible liquid, and the associated decrease in convergence rate, can be circumvented with the incompressible treatment. In addition, a larger time-step can be applied with the incompressible treatment of the liquid, because no shock waves have to be resolved in the liquid. As a result, simply switching to an incompressible treatment of the water drop accelerates the simulation by factor $1.8$, and also adjusting the applied time-step brings an overall acceleration of the simulation of factor $3.9$.

\section{Conclusions}
\label{sec:conclusions}

We have presented a new and promising route to construct numerical algorithms for the simulation of both compressible and incompressible interfacial flows, as well as mixtures thereof, based on a unified thermodynamic closure model, a finite-volume discretisation and a fully-coupled pressure-based algorithm. 
The proposed thermodynamic closure model and treatment of the fluid properties at the interface, in conjunction with the ACID method to couple the interacting bulk phases \cite{Denner2018b} and the Newton linearisation of the governing equations \cite{Denner2018c}, bridge the different numerical requirements for the simulation of incompressible and compressible fluids, and facilitate the simulation of compressible-incompressible interfacial flows. 
If, however, only incompressible fluids or only compressible fluids are considered, the presented algorithm is equivalent to previously proposed pressure-based algorithms for incompressible interfacial flows \cite{Denner2014a} or compressible interfacial flows \cite{Denner2018b}, respectively. 

The proposed algorithm has been successfully validated using five representative test-cases, each featuring a combination of a compressible fluid and an incompressible fluid separated by an interface: a bubble with surface tension in equilibrium, the viscous damping of capillary waves, the reflection of an acoustic wave at a gas-liquid interface, the pressure-driven collapse of a bubble and the shock interaction with a water drop. 
In particular, the presented simulations of interactions of an acoustic wave with a compressible-incompressible interface and of a high-Mach compressible flow with an incompressible fluid ({\em i.e.}~the shock-drop interaction) are unique capabilities that have not been demonstrated in the literature before. In addition, treating liquids as incompressible, while still resolving acoustic effects in the gas phase, has been shown to yield substantial performance benefits.

For gas-liquid flows in general, the strength of the proposed algorithm lies in the reliable prediction of compressible effects in the gas phase, without paying the additional cost of resolving marginal physical effects in the liquid phase, if the application allows this. For instance, acoustic waves in the gas phase are known to promote interfacial instabilities \cite{Zhou1992}, which however have a negligible influence on the behaviour of the liquid phase in subsonic flows; especially, the compressibility of the liquid does not influence the acoustics in the gas phase, as shown by the presented results. With the presented algorithm the simplification of an incompressible fluid cannot only be invoked for low-Mach flows with respect to the gas phase, as considered in previous studies \cite{Billaud2011, Caltagirone2011, Yamamoto2018}, but practically for any flow velocity. While a compressible liquid is not a valid simplification for large liquid Mach numbers, for subsonic gas flows, for instance in subsonic fuel injection and spray atomisation processes, the presented results suggest an incompressible liquid to be a reasonable assumption.

\section*{Acknowledgements}
This research was funded by the Deutsche Forschungsgemeinschaft (DFG, German Research Foundation), grant numbers 420239128 and 447633787.


\end{document}